%% file: main.tex
\documentclass[usenatbib]{mnras}

\usepackage{newtxtext,newtxmath}

\usepackage[T1]{fontenc}
\usepackage{ae,aecompl}

\usepackage{graphicx}	
\usepackage{amsmath}	
\usepackage{amssymb}	
\usepackage{longtable}
\usepackage{CJKutf8}
\usepackage{multicol,multirow}

\newcommand{\degree}{$^\circ$}
\newcommand{\dmunit}{$\mathrm{pc\ cm^{-3}}$}

\title[ASKAP FRB Scattering]{A population analysis of pulse broadening in ASKAP Fast Radio Bursts}

\author[H. Qiu et al.]{Hao Qiu (\begin{CJK*}{UTF8}{gbsn}邱昊\end{CJK*}),$^{1,2}$\thanks{E-mail: hqiu0129@uni.sydney.edu.au}
Ryan M. Shannon,$^{3}$
Wael Farah,$^{3}$
Jean-Pierre Macquart,$^{4,\thanks{Deceased}}$
\newauthor Adam T. Deller,$^{3}$
Keith W. Bannister,$^{2}$ 
Clancy W. James,$^{4}$
Chris Flynn,$^{3}$ 
\newauthor Cherie K. Day,$^{3,2}$ Shivani Bhandari$^{2}$ and Tara Murphy$^{1}$
\\
$^{1}$Sydney Institute for Astronomy, School of Physics, University of Sydney, NSW 2006, Australia\\
$^{2}$CSIRO Astronomy and Space Science, PO Box 76, NSW 1710, Australia\\
$^{3}$Centre for Astrophysics and Supercomputing, Swinburne University of Technology, Mail H30, PO Box 218, VIC 3122, Australia\\
$^{4}$ICRAR/Curtin University, Curtin Institute of Radio Astronomy, WA 6845, Australia
}

\date{Accepted 2020 June 29. Received 2020 June 25; in original form 2020 May 11}

\pubyear{2019}

\begin{document}
\label{firstpage}
\pagerange{\pageref{firstpage}--\pageref{lastpage}}
\maketitle

\begin{abstract}
The pulse morphology of fast radio bursts (FRBs) provides key information
in both understanding progenitor physics and the plasma medium through which the burst propagates.
We present a study of the profiles of 33 bright FRBs detected by the Australian Square Kilometre Array Pathfinder. 
We identify seven FRBs with measureable intrinsic pulse widths, including two FRBs that have been seen to repeat.
In our modest sample we see no evidence for bimodality in the pulse width distribution.
We also identify five FRBs with evidence of millisecond timescale 
pulse broadening caused by scattering in inhomogeneous plasma.
We find no evidence for a relationship between pulse broadening and extragalactic dispersion measure.
The scattering could be either caused by extreme turbulence in the host galaxy
or chance propagation through foreground galaxies.
With future high time resolution observations and detailed study of host galaxy properties we may be able to probe line-of-sight turbulence on gigaparsec scales.

\end{abstract}

\begin{keywords}
fast fadio bursts -- methods: data analysis  --  galaxies: ISM -- intergalactic medium
\end{keywords}



\section{Introduction}
Fast Radio Bursts (FRBs) are a class of extragalactic radio transient which have a durations of the order of microseconds to tens of milliseconds. 
While more than 90 FRBs have been published to date
\citep{2016PASA...33...45P}, their origin is yet to be determined.
The bursts have been detected and studied at frequencies ranging from 300 MHz to 8 GHz \citep{2007Sci...318..777L,2013Sci...341...53T,2018ApJ...863....2G,2018Natur.562..386S,2019Natur.566..230C,2020arXiv200312748P} using both single dish telescopes such as the Parkes 64m radio telscope, and localised to host galaxies using interferometer arrays such as the Very Large Array (VLA; \citealt{2017Natur.541...58C}),
the Australian Square Kilometre Array Pathfinder (ASKAP; \citealt{2017ApJ...841L..12B}) and
the Deep Synoptic Array (DSA; \citealt{2019Natur.572..352R}).
A key feature of FRBs is that the dispersion measures (DMs) of the bursts
exceed the line of sight DM contribution from the Milky Way, and that there is a strong correlation between this excess dispersion measure and source distances \citep{2018Natur.562..386S,2020Natur.581..391M}.  

Analysis of the FRB population has shown significant diversity in pulse properties such as
fluence, DM, spectra and temporal structures including scattering broadening \citep{2019MNRAS.482.1966R}.
Attempts have been made to infer the characteristics of the FRB 
population by studying the distribution of FRB properties such as fluence \citep{2016MNRAS.461..984O}, 
detection rates \citep{2019MNRAS.487.5753C}, DM, and scattering broadening 
\citep{2019MNRAS.482.1966R}. 
The distribution of these properties can be used to characterise the population and potentially identify sub-populations.

Some FRBs have been observed to repeat \citep{2016Natur.531..202S,2019Natur.566..235C}
Using a sample of containing 70 FRBs not yet seen to repeat, \citet{2019NatAs...3..928R} estimated a volumetric rate of $\mathrm{10^5\ Gpc^{-3}\ yr^{-1}}$.  
The incompatibility between this value and the occurrence of potential cataclysmic progenitors (when considering additional constraints based on potential host galaxy contributions to the FRB DM)
led \citet{2019NatAs...3..928R} to conclude that at least a fraction of apparently one-off FRBs must exhibit repeat activity.

It remains unclear whether
repeating and one-off bursts have different origins.
However, many of the repeating bursts have been observed to display 
distinctive pulse profile features such as frequency drifting 
\citep{2019ApJ...876L..23H} and wider pulse profiles 
compared to non-repeating FRBs \citep{2019ApJ...885L..24C}. 
Other properties of the FRBs such as rotation measure (RM), 
host galaxy type and local environment 
may also help understand the connection between repeating and non-repeating events 
\cite[]{2016Natur.531..202S,2019ApJ...887L..30K,2020ApJ...895L..37B}.

Propagation effects such as pulse broadening caused by scattering and spectral
modulation can be used to probe dense plasma
in interstellar medium (ISM), circumgalactic medium (CGM) and the intergalactic medium (IGM) along the line of sight
\citep{2019ApJ...872L..19M,2016arXiv160505890C,2014ApJ...797...71Z}. 
While the Milky Way can contribute to these effects, many FRBs are detected at higher Galactic latitudes where the dispersion measure contribution from the Milky Way is small, and the scattering
caused by the Milky Way ISM is weak \citep{2019ARA&A..57..417C}.
Additionally the decorrelation scale of the scintillation seen in many FRBs is inconsistent
with that arising from the Milky Way ISM alone \citep{2018MNRAS.478.1209F,2019ApJ...872L..19M}.
Consequently, most temporal broadening in FRBs 
is expected to be caused by ionised media beyond the Milky Way. 
However, it is unclear where the plasma causing most of the scattering is located.
It has been suggested that the scatter broadening can characterise the local environments of the FRBs \citep{2016ApJ...818...19K,2016arXiv160505890C,2019ARA&A..57..417C}. It is also possible that the pulse broadening originates from the IGM and the CGM of intervening galaxies \citep{2019Sci...366..231P,2020ApJ...891L..38C}.
A correlation between the scattering time, $\tau$, and the extragalactic DM of the FRB population might be present in this latter case.

The Australian Square Kilometre Array Pathfinder (ASKAP,
\citealp{2016PASA...33...42M}) has detected 33 bright FRBs during 2017-2019
\citep{2018Natur.562..386S,2019Sci...365..565B,2019MNRAS.486..166Q,2019MNRAS.486...70B,2019MNRAS.490....1A}, providing a relatively homogeneous collection of bursts with high signal to noise and sufficient bandwidth to explore the pulse morphology.

In this work, we use Bayesian methodology to model the ASKAP FRB pulse profiles.
{The methods, models and analysis setup are discussed in Section \ref{sec:methods}.
We report the evidence of scatter broadening, scattering time and the intrinsic 
pulse widths from the FRBs in Section \ref{sec:analysis} with population analysis using Bayesian inference.  
The discussions of the results are then presented in Section \ref{sec:discussion}.}


\section{Data Collection}
\subsection{ASKAP FRBs}
Data for 33 FRBs detected by ASKAP are used in this analysis,
including both FRBs detected in single dish fly's-eye observations
described in \citet{2017ApJ...841L..12B,2018Natur.562..386S} and those detected during interferometric observations \citep{2019Sci...365..565B,2019Sci...366..231P,2020Natur.581..391M,2020ApJ...895L..37B}.
The Commensal Real-time ASKAP Fast Transients 
(CRAFT, \citealp{2010PASA...27..272M}) pipeline produces two data products during observations: low time resolution dynamic spectra of total intensity generated for candidate detection, and per-antenna raw voltages that can be saved from short buffers after candidate detection and used to produce high resolution dynamic spectra. Low time resolution data are available for all FRBs, while the voltage capture data are available for FRBs detected in interferometric observations. 

The low time resolution candidate pipeline creates dynamic spectra with 
approximately 1 ms time resolution for 
each of the 36 beams in the field of observation. 
If multiple antennas are simultaneously used, the pipeline generates the dynamic spectra using the incoherent sum of each beam pointing.
The low resolution data has a total bandwidth of 336 MHz in 
1 MHz bandwidth channels.
{The ASKAP FRB modes can observe in a $336$~MHz band in the frequency range 700--1800 MHz.
The time resolution and observing frequency for the dynamic spectra have been
slightly adjusted over the course of the project. The time resolution 
has been changed between 0.86--1.73 ms as shown in Table \ref{tab:obs_spec}.}

The voltage capture system is used in FRB detections after August 2018
for interferometric localisation \citep{2019Sci...365..565B} and 
the data are also used for developing high time resolution dynamic spectra of the FRBs \citep{2020ApJ...891L..38C,2020arXiv200513162D}.

The FRBs detected prior to the commencement of localisation searches only have data from the detection pipeline,  and were limited to a  time resolution 
of 1.265 ms. The localised FRBs have both the detection pipeline low resolution data and a reconstructed high time resolution time series after data reduction.
For the purpose of this paper, we choose to use the low resolution ASKAP dynamic spectra generated in the commensal FRB detection pipeline of all FRBs as a near-uniform sample for analysis.
We check the validity of the low resolution data compared with high resolution data in Section \ref{sec:verification}.
{The dynamic spectrum of all FRBs in this paper can be found in the following publications: \citet{2018Natur.562..386S}, \citet{2019ApJ...872L..19M},
\citet{2019Sci...365..565B},\citet{2019Sci...366..231P},
\citet{2020ApJ...891L..38C},
\citet{2020ApJ...895L..37B} and \citet{2020arXiv200513162D}.}

\begin{table}
	\centering
	\caption{Observation Specifications of the FRB observations. We specify the period of time, centre frequency and sampling time resolution ($\tau_{\rm samp}$).}
	\label{tab:obs_spec}
	\begin{tabular}{lll} 
		\hline
	Time Period &Centre Frequency & $\tau_{\rm samp}$  \\
	&(MHz) &(ms)\\
		\hline

2017 Jan -- 2018 Jul$^{a}$& 1297.5 &1.265\\
2018 Aug -- 2018 Dec &1298.5 &0.864\\
2019 Jan&1272.5 & 0.864\\
2019 Jun --2019 July&1272.5 &1.728\\
		\hline
	\end{tabular}

\end{table}
        

\subsection{Data preparation}

All FRBs were first dedispersed based on previously reported values \citep{2018Natur.562..386S}. 
Frequency channels are individually inspected for radio-frequency interference (RFI) and contaminated channels are replaced with Gaussian noise with zero mean and unit variance. 

The data analysis of each burst uses cutouts starting 15 time samples before the burst and extending to 45 time samples after. This allows us to measure pulse broadening up to 30 ms (much longer than observed in any of the bursts).

The noise level for each channel is measured from data before the cutout and then used
to rescale each channel to have uniform variance in the signal-free regions. This helps to verify the Bayesian analysis method in the following section.
For the multi-frequency analysis, we average and scale the data into 8-subband spectra.

\section{Methods and modelling}
\label{sec:methods}
\subsection{Nested sampling method and Bayesian statistics}
We use a Bayesian approach for the multifrequency modelling to determine the best model parameters, 
similar to the technique employed in \citet{2019MNRAS.482.1966R} for FRBs detected by the Parkes radio telescope. 
Bayesian methods account for the covariances between parameters and provides robust parameter estimation.
We can use this method to verify the rescaling, through characterisation of the noise in the cutout. We use a Gaussian likelihood function for our maximum likelihood estimation:
\begin{equation}
\mathcal{L}(x_i|\mu,\sigma)=\frac{1}{\sqrt{2\pi\sigma^2}}\rm {exp}\left[\frac{-(x_i-\mu)^2}{2\sigma^2}\right].
\label{eq:Likelihood}
\end{equation}

We use nested sampling techniques implemented in the python packages
\textsc{Dynesty} \citep{2020MNRAS.493.3132S} and \textsc{Bilby} \citep{2019ApJS..241...27A}. 
Nested sampling provides an efficient way to calculate the marginal likelihood, which can be used for model comparison. 
The Bayes factor is used to measure which model is preferred:
\begin{equation}
B_{ij}=\frac{p(D|M_i I)}{p(D|M_j I)}=\frac{L(M_i)}{L(M_j)}
,
\label{eq:Bayes}
\end{equation}
where $D$ is the data, $I$ is the prior information, $M_i$ and $M_j$ are the two different models. 
$p(D|M I)$ is the probability of obtaining the observed data with given model and prior. $L(M)$ is the likelihood function or the Bayesian model evidence.


We also apply hierarchical Bayesian inference to search for correlations
in the population.
This allows us search to include marginal detections and non-detections in the analysis. 
The total hyper-parameter likelihood function is:
\begin{equation}
\mathcal{L}_{\mathrm{tot}}(\Vec{d}|\Lambda)=
\prod_{i}^{N_{\mathrm{FRB}}}\frac{\mathcal{Z}_\mathrm{M}(d_i)}{n_i}
\sum_{j}^{n_i}\frac{\pi(\theta^k_i|\Lambda)}{\pi(\theta^k_i|\mathrm{M})}
,
\label{eq:hyper}
\end{equation}
where $\mathcal{Z}_\mathrm{M}(d_i)$ is the evidence for fitting each FRB, 
${n_i}$ is the number of posterior samples and
${\pi(\theta^k_i|\mathrm{M})}$ is the default prior for that posterior \citep{2019PASA...36...10T}.

\subsection{Pulse profile modelling}

We start with a simple Gaussian model for a dedispersed FRB pulse profile. 
The millisecond timescale resolution of the data means we will not 
resolve any microstructure within FRBs as observed in \citet{2018MNRAS.478.1209F}. 
To correctly model FRBs that are temporally unresolved in coarse time resolution, pulses are constructed in an array with a sampling interval 10 times higher than the filterbank data
and then averaged in time to the final temporal resolution of the data.
The base model for the pulse at subband $i$ is a Gaussian function: 
\begin{equation}
f_i(t)=\frac{\rm{A_i}}{\sqrt{2\pi\sigma^2}} \exp \left[\frac{-(t-t_0-t_i)^2}{\sigma^2} \right],
\label{eq:gaus}
\end{equation}
where $\rm{A_i}$ is the amplitude coefficient, $t_0$ is the time reference of the burst at the top band frequency and $\sigma$ is the apparent width of the pulse at this frequency. 
The parameter $t_i$ compensates for pulse shift compared to the top band time reference induced by previously unmodelled dispersion, characterised by the dispersion measure offset ($\mathrm{DM_{off}}$)
from the original detection value DM.
The total dispersion measure of the burst is defined as $\mathrm{DM_{tot}=DM+DM}_\mathrm{off}$ where DM is the reported value from the pipeline used to dedisperse the pulse.
We estimate the time delay correction in each channel centred at frequency $\nu_i$ using the dispersion relation:
\begin{equation}
t_i=4.15\ \mathrm{ms\ DM}_{\mathrm{off}}(\nu_{\mathrm{top}}^{-2}-\nu_i^{-2}),
\label{eq:tidm}
\end{equation}
where $\nu_{\mathrm{top}}$ is the top band frequency of the data.

The duration (width) of the pulse, $\sigma$, consists of two terms: the intrinsic width($\sigma_i$)  and the intrachannel dispersion smearing ($\sigma_{\mathrm{DM}}$),
$\sigma=(\sigma_i^2+\sigma_{\mathrm{DM}}^2)^{1/2}$. The smearing is
caused by intra-channel dispersion for which the half width is given by:
\begin{equation}
\sigma_{\mathrm{DM}}=(4.15\times 10^{-3} \rm{ms} )\ \mathrm{B\ {DM_{tot}}}\  \nu^{-3} .
\label{eq:smear}
\end{equation}
$B$ is the bandwidth of the channel in MHz
and $\nu$ is the centre frequency of the channel in GHz.
We note that the ASKAP {data are produced by a polyphase filter, oversampled by a factor of 32/27,  so the effective bandwidth is marginally higher than 1~MHz and adjacent channels are slightly correlated.} The wider effective bandwidth is approximately 
1.1~MHz \citep{tuthill2015compensating} which increases the dispersion smearing by  $\sim$ 0.05 ms (for DM $\sim$ 300 \dmunit) at 1.3 GHz.

To account for pulse broadening caused by multipath scattering,
we apply an exponential decay function that is convolved with the Gaussian expressed in Equation~\ref{eq:gaus},
to create a scattered-burst profile.
The convolution kernel to create pulse broadening is described as:
\begin{equation}
f_i(t)=\begin{cases}
    \exp\left[-\frac{(t-t_s)}{\tau({\nu_i}/{\nu_{1.3}})^{-\alpha}}\right], & (t\geq t_s),\\
    0 ,         & (t< t_s).
\end{cases}
\label{eq:exponent}
\end{equation}
we define $t_s$ as the incident time of the scattering.
The scattering broadening timescale $\tau$ is reported at 1.3 GHz and is scaled to 
individual channels at frequency $\nu_i$, using a power law exponent spectral index $\alpha$. 
For most bursts we assume a fixed index of $\alpha=4$ based on pulsar scattering observations through the ISM.
However, for the bright and significantly scattered FRB~180110 we fit with the spectral index $\alpha$ as a free parameter.

\subsection{Analysis setup}
We compare the Bayesian evidences to identify pulse broadening in the FRBs.
We apply two models as described in the previous section: (a) a simple Gaussian model (SG) and (b) a broadened Gaussian model (BG) {which is formed by the Gaussian model convolved with the exponential decay function. }
We use the logarithmic Bayes factor to distinguish the preferred model.
We specify the logarithmic Bayes factor in following sections as the difference between model (b) and model (a):
\begin{equation}
{\mathrm{log(B_{ba})}=\mathrm{logL(M_b)-logL(M_a)}=\Delta {\mathrm{log E}},}
\label{eq:deltaE}
\end{equation}
{$\Delta {\mathrm{log E}}$ is the difference between the marginal likelihood or Bayesian evidence of the two models.} A positive Bayes factor indicates evidence for pulse broadening. {We use the Jeffreys' scale \citep{Jeffreys61,2008ConPh..49...71T} as a interpretation of the evidence: $ {\mathrm{log_{10} B}}\leq1$ as weak inconclusive evidence ($\mathrm{P} < 75\%$), $1< {\mathrm{log_{10} B}}<10$ as modest substantial evidence\footnote{The probability at $\mathrm{log_{10} B=5}$ is\ $\mathrm{P} = 99.3\%$}, $ {\mathrm{log_{10} B}}\geq10$ is considered as strong evidence.}

The nested sampling in this work uses 300 live sampler points with the termination condition set at dLog$E$<0.1, 
where dLog$E$ is the differential of the logarithimic marginal likelihood function. The marginal likelihood for each model is then used to calculate the Bayes factor as an indicator of confidence in the model. 


To assess the effectiveness of measuring 
scattering timescales and pulse widths, 
we simulated FRBs with DMs of 
100--1000 $\mathrm{pc\ cm}^{-3}$ spanning our observed range.
The FRBs were created with an intrinsic width of 1.3 ms, 
3 different pulse broadening times 
(0, 1 and 5 ms with $\alpha=4$) and 2 different values for signal-to-noise ratio (S/N) 
in a single sub-band (5 and 30).
A set of 10 FRBs were simulated for each set of the parameters with random noise realisations.

\begin{figure}
    \centering
    	\includegraphics[trim={0 0.0cm 0 0.0cm },clip,width=\columnwidth]{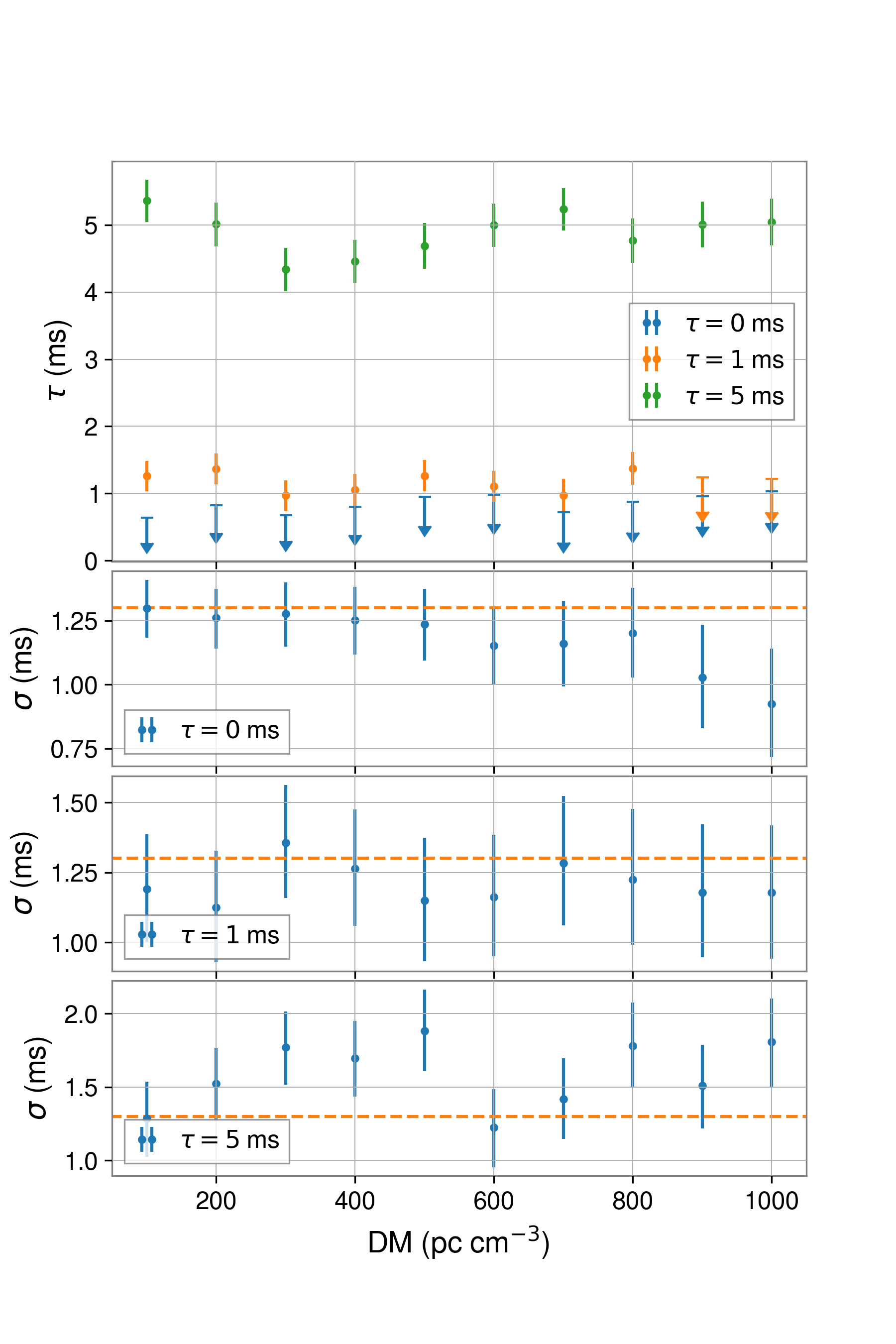}
    \caption{{Results of average measured pulse broadening ($\tau$) and  intrinsic width ($\sigma$) from simulated FRBs with a subband S/N of 5 and intrinsic width of 1.3ms.,
    upper limits are plotted for indefinitive results. Each data point represents the average result of 10 simulated bursts at that DM and broadening setting with random noise. The orange dashed line indicates the width of the simulated pulse.
    }}
    \label{fig:simulation}

    \label{fig:width_sim}
\end{figure}

The fitting results from simulated FRBs are not heavily affected by dispersion smearing, and the parameters for the high S/N pulses are measured reliably.
We plot the average scatter broadening measured from low S/N simulation data over DM in Figure \ref{fig:simulation}.
When the pulses are weaker, we are still able to detect scattering. 
However it becomes more difficult to measure the intrinsic pulse width, especially at high dispersion measure (see Figure \ref{fig:width_sim}) it will be harder to separate the intrinsic width from the smearing width. 
For larger scattering times, the pulse width measurement is scattered and highly dependent on the random noise generated for the simulation.
In this work, only 1 FRB (FRB 170428) out of 33 has a DM higher than 800 \dmunit which may be affected by this issue,the FRB does not show scattering and we are only able to provide an upper limit on the pulse width where smearing width.


\section{Pulse Analysis}
\label{sec:analysis}
\subsection{Fitting results and evidence of scattering}
\label{sec:results}

The results from the profile fitting are presented in Table \ref{tab:fit}. 
New measurements of intrinsic pulse width and updated DM values of each FRB from the preferred of the two models compared are listed in the table. 
We also estimate the extragalactic DM contribution by subtracting the Galactic DM contribution given by the NE2001 electron density model\citep{2003astro.ph..1598C}.
The Bayes factor is calculated for each FRB to measure the probability of scatter broadening and select the best fitting model.
Upper limits are provided to those that have minimal pulse broadening in the posterior distribution.

Five FRBs have a positive Bayes factor, indicating evidence for scatter broadening. FRB 180110 has strong evidence shown in the model comparison \citep[${\mathrm{log_{10} B}}$> 10, ][]{Jeffreys61,2008ConPh..49...71T}; 
FRB 180119 and FRB 180130 have modest evidence (1<${\mathrm{log_{10} B}}$<10); 
FRB 180324 and FRB 180525 have weak evidence (${\mathrm{log_{10} B}}$<1, P<75\%) for scatter broadening.
We are limited by the time and frequency resolution of our data, as discussed in Section 2 and 3. 
High time resolution data of localised FRBs could be used to measure temporal broadening due to scattering at shorter time scales.

\begin{table*}
\centering
\bgroup
\def\arraystretch{1.2}
	\centering
	\caption{{ASKAP FRB pulse profile properties. 
	The following parameters are presented in this table: total DM ($\mathrm{DM_{tot}}$), extragalactic DM contribution ($\mathrm{DM_{Extragalactic}}$), intrinsic pulse width from Gaussian model ($\sigma_{\mathrm{SG}}$), intrinsic pulse width from scattering model ($\sigma_{\mathrm{BG}}$), scattering time at 1.3 GHz ($\tau_{\rm{1.3 GHz}}^b$) and logarithimic Bayes Factor (log$_{10}$B).  
	The Bayes factor between models including and excluding scatter broadening which is used to select a preferred model.  The preferred models are highlighted in Italic font.}
	The DM of the burst is taken from the preferred model. The scattering timescale $\tau$ is recorded from the posterior of the scattering model, for FRBs that show no evidence of scatter broadening we provide upper limit measurements.
	Uncertainties are calculated at 1-$\sigma$ confidence.}
	\label{tab:fit}
	\begin{tabular}{llclllr} 
	
	\hline
FRB & DM$_{\rm tot}^a$
&DM$_{\mathrm{Extragalactic}}$& $\sigma_{\mathrm{SG}}^{a,b}$& $\sigma_{\mathrm{BG}}^{a,b}$ & $\tau_{\rm{1.3 GHz}}^b$ & log$_{10}$B \\
&(pc cm$^{-3}$)&(pc cm$^{-3}$)&(ms)&(ms)&(ms)\\
\hline
\input{scatfit.tex}

\hline

\multicolumn{5}{l}{a. Based on the results from best model}\\
\multicolumn{5}{l}{b. Upper limits are calculated at 90\% confidence}\\
\multicolumn{5}{l}{$\dagger$\ FRBs with repetition activity detected}\\
	\end{tabular}
\egroup
\end{table*}





\input{section4.tex}

\subsection{Result verification}
\label{sec:verification}
\begin{table}
\centering
\bgroup
\def\arraystretch{1.2}
	\centering
	\caption{{Comparison of the intrinsic pulse width ($\sigma$) and scattering time ($\tau$) measurements from low resolution filterbank data (FB) in this work and high time resolution data (HTR) measured in \citet{2020ApJ...891L..38C} and
	\citet{2020arXiv200513162D}. }}
	\label{tab:compare}
	\begin{tabular}{llllll} 
	
	\hline
FRB &
 $\sigma_{\mathrm{FB}}$& $\sigma_{\mathrm{HTR}}$ & $\tau_{\rm{FB}}$ & $\tau_{\rm{HTR}}$\\
&(ms)&(ms)&(ms)&(ms)\\
\hline
180924& $<0.63$&$0.09\pm0.04$&$<0.72$&$0.68	\pm 0.03$\\
181112&$<0.58$& $0.016\pm0.001$&$<0.55$& $0.021\pm0.001$\\
190102& $<0.62$&  $0.053\pm0.002$  & $<0.55$  &
$0.041	^{+0.002}_{-0.003}$\\
190608& $2.22^{+0.63}_{-0.64}$ & $1.1\pm0.2$  &  $<3.58$ & $3.3\pm 0.2$  \\
190611 & $1.03^{+0.61}_{-0.63}$& $0.09\pm0.02^\dagger$  & $<1.45$  & $0.18\pm 0.02$\\
\hline
\multicolumn{5}{l}{$\dagger$. Width of first pulse component}\\
\multicolumn{5}{l}{$\dagger\dagger$. No attempt was made to fit the complex time-domain}\\
\multicolumn{5}{l}{\space \space \space structure of 190711 in \citet{2020arXiv200513162D}}\\
\end{tabular}
\egroup
\end{table}
In addition to simulations we have two other checks of our methodology. 
We show the FRB widths as a function of DM in Figure \ref{fig:width_dm}. There is no sign of correlation between DM and width, which confirms that the method is able to separate the dispersion smearing component of the pulse.

For six of the bursts reported here 
(FRBs 180924, 181112, 190102, 190608, 190611 and 190711),
we have access to high time resolution data against which to compare our results 
\citep{2020ApJ...891L..38C,2020arXiv200513162D}.
The high time resolution data has higher signal to noise ratio both because it is not subject to dispersion smearing, and because it is produced from an array coherent data product as opposed to incoherent summation in the detection pipeline.
For FRB 180924, the scattering time in the high time resolution data is measured to be $\tau= 0.68\pm0.03$ ms, shorter than the time resolution of the data, and consistent with our upper limits.
In the high time resolution data, FRB 190608 shows evidence for a wide pulse width and pulse broadening of $\tau= 3.3\pm0.2$ ms \citep{2020arXiv200513162D}. We identify the pulse in the low resolution data to be resolved, and place a limit on the scatter broadening consistent with the upper limits. 

In the high time resolution data, FRBs 181112, 190102, 190611 and 190711 show multiple components with less than 1~ms separation. 
We do not fit for the multiple components in the detection pipeline data due to dispersion smearing and low time resolution.

For FRBs 181112 and 190102, the signal is dominated by one narrow component and the upper limits from the low resolution data are consistent with the pulse.
The time separation between the peak components in the high time resolution data of FRB 190611 is below the time resolution of the detection pipeline data. Our pulse width measurement is consistent with the pulse separation of the 2 pulses.
FRB 190711 shows complex structure in high time resolution data during a $\sim 10$ ms duration with no evidence for pulse broadening.
This complex structure is not visible in the detection pipeline data because the complex pulse structure is partially below the pipeline data sensitivity. Our measurements are consistent with the primary component width at high time resolution.

{
In summary, the low time resolution data gives results as expected. We are not able to identify faint or extremely short timescale secondary components in low time resolution data. 
However, we are able to identify large pulse broadening time scales and wide FRBs.
These properties may help with the study of dense local media around the FRBs and identify possible repeating FRBs \citep{2020arXiv200311930C}.}

\subsection{Relation between Galactic DM and scatter broadening}
\label{sec:gal_dm_scatter}
The distribution of pulse broadening and total DM for our sample of FRBs is shown in Figure \ref{fig:scat_total}. 
The FRBs for which there is no evidence for scatter broadening are plotted as 90\% upper limits. 
We find no obvious correlation between the DM and the pulse broadening. 
We separate the Galactic DM and extragalactic DM
contribution to constrain and identify the host medium
responsible for temporal broadening.

In Figure \ref{fig:scat_ism} we show the distribution of broadening times compared to the predicted line of sight Galactic DM contributions from the NE2001 model.
Most  FRBs detected in this sample 
are located at high Galactic latitudes to avoid large Galactic DM contributions \citep{2018Natur.562..386S}. 
This in most cases leaves $\mathrm{< 50}$ \dmunit\ Galactic DM contribution to the line of sight. 

The Galactic scattering measure (SM) of all high Galactic latitude detections 
are $< 10^{-3}\ \mathrm {kpc\ m^{-20/3}}$,
from which the estimate temporal broadening from the Galactic scattering measure can be calculated as follows \citep{2002astro.ph..7156C}:
\begin{equation}
\tau_d=1.10\ \mathrm{ms}\ \mathrm{SM}_{\tau}^{6/5} \nu^{-22/5} D,
\label{eq:SM_scat}
\end{equation}
where $D$ is the distance to the scattering screen in kpc and $\nu$ is the observed frequency in GHz. 
The SM of these FRBs limits the scatter broadening provided by the Milky Way to
$\tau<\mathrm{2 \mu s} $.

The exception is FRB 180430 \citep{2019MNRAS.486..166Q}, which was detected in the Galactic Plane. However, the FRB was observed at a Galactic anticentre direction with a low predicted Galactic $\mathrm{SM}\sim 10^{-2.44}\ \mathrm {kpc\ m^{-20/3}}$. The burst also shows no evidence of pulse broadening.


\subsection{Scattering index of FRB 180110}
\label{sec:measure180110}
FRB 180110 is an FRB with high S/N and a significant scattering tail.
We further investigate the frequency dependence of the pulse by using the scatter broadening model with an unconstrained scattering spectral index.
We display the fitting posterior distribution in Figure \ref{fig:post_180110} and sub-band residuals of the model in Figure \ref{fig:residual_180110}. 
The result shows a broadening time scale of $\mathrm{5.9 \pm 0.4ms}$ at 1 GHz, 
with the pulse broadening is proportional to $\nu^{-\alpha}$ with a scaling index of $\alpha=3.7^{+0.9}_{-0.9}$.

This frequency dependence is consistent with the spectral index measured from pulsars scattered by the ISM ($\alpha\sim 4$, \citealt{2004ApJ...605..759B}).
However, FRB 180110 was detected at high Galactic latitude of $|b| \sim 50$ \degree, and as discussed above, the Galactic SM contribution from simulated models cannot produce scattering of the observed magnitude. {No known HII region, which may provided unexpected density flucuations, is located within the localisation region of FRB 180110. 
The lack of evidence for turbulence indicates that the scattering is likely to be caused by propagation through plasma outside the Milky Way, either in the host galaxy or in an intervening galaxy or galaxy halo.}

\subsection{Hierarchical Bayesian inference of the extragalactic DM-scatter relation}
\label{sec:egDM-tau}
We then search for correlation between the extragalctic DM (DM$_\mathrm{EG}$) and broadening time of the FRBs. 
We use the posterior distribution of $\tau$ for each burst to apply Bayesian inference on the extragalactic DM (DM$_\mathrm{EG}$) and scattering ($\tau$) relation of the FRBs. 
This allows us to include upper limits from the posterior distribution and consider the uncertain scatter .

To distinguish whether the small number of FRBs with scattering broadening are special cases, we search for the correlation on 2 sets of data: (1) all FRBs and (2) FRBs with pulse broadening.
{For FRBs with higher time resolution data, we use the scattering time measured in \citet{2020arXiv200513162D} to reduce the uncertainty of scattering.}

We test two models to see if there is a relation between DM$_\mathrm{EG}$ and $\tau$. 
The first model is a flat log\,$\tau$=log\,A,  i.e., there is no detected relation
between DM and $\tau$. 
We compare this with a log linear model which represents a plausible relation between the DM$_\mathrm{EG}$ and $\tau$ in the form of $\tau$=A DM$^\alpha$. 
We plot the data and models in Figure \ref{fig:scat_dm}.
We also show the empirical DM-scattering relation derived from the pulsars
located in the
inner Galactic disk \citet{2004ApJ...605..759B} as a reference to compare with  the extragalactic estimate relation. Due to the few number of confirmed scattering cases, we do not consider the more complicated (superexponential) relationships applied ot pulsars and used in \citet{1991ApJ...376..123C},\citet{2004ApJ...605..759B} 
and \citet{2019ARA&A..57..417C}.


{Both model comparisons using all FRBs and FRBs with confirmed scattering show no firm evidence to support the log linear model (Log\,B $< 1$, P < 75\%).} 
The FRBs are inconsistent with the scaled Galactic pulsar relation in \citet{2019ARA&A..57..417C}. 
The number of upper limit measurements at low $\mathrm{DM_{EG}}$ heavily constrains the slope of the fit, which indicates it is unlikely there is a proportional relation between DM$_\mathrm{EG}$ and $\tau$ for the detected ASKAP FRBs.

In summary, the model comparison does not favour the log linear 
model. Given our small data set of samples, we do not find evidence for a relation between
$\rm{DM_{EG}}$ and $\tau$.

\section{Discussion}
\label{sec:discussion}

{As most of the bursts are at high Galactic latitude, it is unlikely that the Milky Way ISM can produce significant ($\gtrsim2\mu$s) pulse broadening. There are therefore} two favoured locations for the scatter broadening found in the FRBs in this survey:  the intergalactic medium or the host galaxy. 

\subsection{Scattering from host environment}

The temporal broadening of ASKAP FRBs is unlikely to be caused by Milky Way-like ISM in the host galaxy.
In this case, a significant amount of host DM  (>200 \dmunit) would be required to cause millisecond timescale broadening as shown by the $\rm{DM-\tau}$ relation in \citet{2004ApJ...605..759B} also shown in 
Figure \ref{fig:scat_dm}.
 
The observed dispersion measure depends strongly on the viewing inclination angle of the host galaxy
as shown from simulations in \citet{2015RAA....15.1629X}. 
In most cases (90\%) the host DM contribution would be $\mathrm{DM_{Host}}< 100$ \dmunit. 
It is possible that the scattered FRBs originated in galaxies aligned near to ``edge-on''. 
The maximum DM contribution of a face-on galaxy with a inclination angle  $\theta > 70$\degree would exceed 
100 \dmunit\ near centre regions. 
The rare case where the FRB travels through the entire thin disk of a spiral galaxy may contribute up to $\sim$ 4000 \dmunit, but no published FRB has been detected with a DM > 3000 \dmunit.
Moreover many localised FRBs have been found to originate outside of 
galaxy centres ($> 4$ kpc,
\citealp{2019Sci...365..565B}, \citealp{2020Natur.577..190M}, \citealp{2020arXiv200513158C}) which would reduce the host galaxy contribution.

We cannot separate the host galaxy and IGM contribution to DM for most FRBs in this sample due to a lack of host galaxy localisation and redshift measures.
However observations of the small sample of localised FRBs show a relationship between extragalactic DM and the redshift of the host galaxy \cite[the Macquart relation,][]{2020Natur.581..391M}.
This suggests that for most FRBs the intergalactic medium has significant contribution to the extragalactic DM.

If the pulse broadening originates from the host galaxy, we would not expect to find a
$\mathrm{DM_{EG}}-\tau$ correlation within the population.
Instead, the amount of pulse broadening would be highly dependent on 
the type and inclination angle of the host galaxy instead of $\mathrm{DM_{EG}}$.

However, any evolution of the FRB population with redshift (either in prevalence or their typical environment) could act to induce an apparent dependence on redshift. 
Given the large scatter we see in the $\rm{DM_{EG}-\tau}$ distributions, and weak evidence favouring a dipersino measure and hence redshift dependence, we still consider this scenario as plausible.
\subsection{Extragalactic scatter broadening}



%


It has been observed that an FRB in some cases could pass through the ISM or halo of a foreground galaxy \citep{2019Sci...366..231P}.
In this section we discuss the feasibility of an extragalactic scattering screen as the origin of the broadening, such as from an intervening galaxy halo or galaxy cluster.

The temporal broadening at cosmological distances \citep{2013ApJ...776..125M} due to an extragalactic scattering screen along the line of sight can be modelled as
\begin{equation}
\tau = \frac{D_L D_{LS} \lambda_0}{2\pi c k D_S(1+z_L)r_\mathrm{diff}^2},
\label{eq:leverarm}
\end{equation}
where $D_S$ is the distance to the source, $D_L$ is the distance to the scatter screen and $D_{LS}$ is the distance from the source to the scatter screen. $z_L$ is the redshift of the foreground screen, $r_\mathrm{diff}$ is the diffractive scale and k is $2\pi/ \lambda_0$ where $ \lambda_0$ is the wavelength of the observer frame.

The scattering produced in intervening halos and intergalactic medium is increased by the large lever arm afforded by the Gpc distance scales. 
If we compare  a screen half-way between the burst source and us ($D_L=0.5 D_S$) to one either in the host or in our own galaxy ($D_L=10^{-6}D_S$), for equally turbulent plasma, the intergalactic screen produces broadening times ${2.5\times10^{5}}$ longer.
The scattering strength is quantified by the scattering measure (SM), where the diffractive scale is proportional to SM$^{1/(\beta-2)}$, where $\beta=11/3$ corresponding to Kolmogorov turbulence. 
To achieve the same level of pulse broadening, 
the scattering strength can be much lower for an extragalactic screen compared to Galactic or host-galaxy media, 
and lower than the observed scattering-DM relationship for the Milky Way.

To achieve pulse broadening that we observe ($\tau \sim 1$\ ms), for an extragalactic screen at $z \sim 0.15$, $D_L=0.625\ \mathrm{Gpc}$ and a FRB at $D_S=1.2\ \mathrm{Gpc}$, 
requires  $r_\mathrm{diff} \approx 2.5 \times 10^{10} \mathrm{cm}$, 
which corresponds to SM >
$2.4 \times 10^{13} \mathrm{m^{-17/3}}$ or 
$7.8 \times 10^{-7} \mathrm{kpc\ m^{-20/3}}$. 
This is over four orders of magnitude smaller than the estimated line of sight Galactic SM contribution of the ASKAP FRBs detected at $b=50$\degree. The DM required to provide such SM would also be significantly smaller than the line of sight Galactic DM (<50 \dmunit). 

If the scatter broadening observed in FRBs originates in extragalactic screens, the further away the FRB, the higher the
probability of the pulse traversing a galaxy or galaxy cluster.
In this case, if $\mathrm{DM_{EG}}$ is proportional to the redshift of an FRB, $\tau$ would also potentially
display correlation to $\mathrm{DM_{EG}}$.

However, there is great uncertainty in the number of intersected foreground galaxies due to the stochastic nature of such events. Additionally the variable properties of the foreground galaxy may not be able to cause scatter broadening.
For the observed foreground galaxy halo for FRB 181112 \citep{2019Sci...366..231P,2020ApJ...891L..38C} and foreground fields of FRB190608 \citep{2020arXiv200513157S}, we do not detect any evidence for scatter broadening. Thus it is possible that many dense foreground galaxies haloes might not be sufficiently turbulent or dense to cause scatter broadening.

This subsection shows that extragalactic screen causing scattering is physically plausible but we need a larger localised FRB sample to discuss the likelihood of this scenario to occur. 

\subsection{Pulse width distribution of ASKAP FRBs}
It has recently been noted that  repeating FRBs apparently have longer pulse durations compared to one-off events
\citep{2020ApJ...891L...6F,2016ApJ...833..177S}.
It is unknown whether this is evidence that repeating FRBs have a different origin or that it is
an selection effect such as repeating bursts have a wider beaming angle, which would make repetition more likely to be
detected \cite[][]{2020arXiv200311930C}. 
Many repeating FRBs have been observed to have intrinsic microstructure \citep{2019ApJ...876L..23H,2019Natur.566..235C} 
and the wider width could also be due to the result of poor time resolution.
\begin{figure}
    \centering
	\includegraphics[trim={0 0 0 0 },clip,width=\columnwidth]{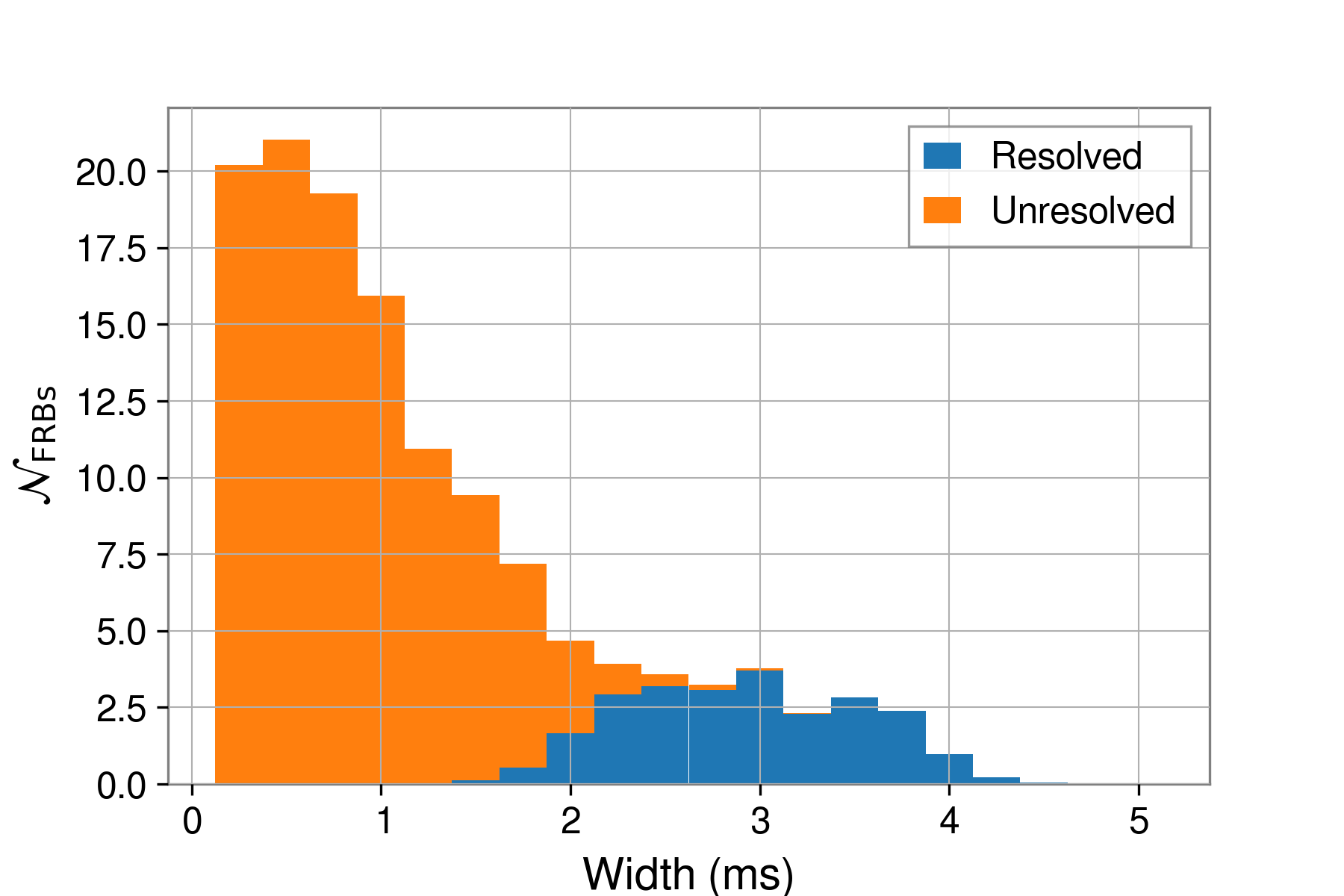}

    \caption{Stacked normalised posterior density of intrinsic pulse widths from the 33 ASKAP FRB sample, unresolved pulses and resolved pulses are separately labelled to display the distribution of resolved pulses}
 \label{fig:stackedwidth}
\end{figure}

In this ASKAP sample, FRBs 170416, 170707, 171019, 180131, 180525, 190608, and 190711 have resolved pulse widths. 
We plot the stacked normalised posterior distribution of the ASKAP FRB widths in Figure \ref{fig:stackedwidth}. The resolved population and non-resolved population in the figure are separately coloured to highlight the distribution of the seven resolved FRBs.
We see no evidence for bimodality in pulse widths in our sample. 
In the resolved sub-sample, we have detected repetition activity from two FRBs.
FRB 171019 \citep{2019ApJ...887L..30K} has been observed to repeat in observations with the Green Bank Telescope. 
Repetition activity has been detected from FRB 190711 (Kumar et al. in prep) with the Parkes radio telescope.
No FRB from the unresolved sub-sample has been detected to repeat yet.


The time resolution of the CHIME search back-end is similar to the ASKAP low resolution data.
A simulated CHIME-detectable repeating FRB population as shown from Figure 2 (right panel) in \citet{2020arXiv200311930C} describes the pulse widths distributed approximately log-normally with a $1\sigma$ range from 1.6 to 45~ms.
The width of the resolved ASKAP pulses are in this range, but with a small sample and only 2 observed to repeat, we cannot test the predictions of \citet{2020arXiv200311930C}.
We also note that the search pipeline for ASKAP FRBs does not search for any pulses wider than 13\,ms, however no FRB has been found with ASKAP that has a total width greater than 7 ms.





\section{Conclusions}
In this work, we analyse the temporal profile of 33 FRBs detected by ASKAP. 
The pulse profile analysis provides updated measurements of DM, pulse intrinsic width and pulse broadening time. 
 

We identify seven bursts with wider intrinsic pulse width. 
The repeating FRB in the ASKAP population both belong to this group of wider pulses. 
The small number of wider pulses in the limited sample of 33 ASKAP FRBs do not show any independent distribution,
which may lead to two separate sub-populations of bursts.

We use Bayesian inference to find evidence for millisecond timescale scattering in five of the bursts.
There is no strong evidence of correlation between DM and scatter broadening.
It is unlikely that the host galaxy could produce the observed level of scattering from a Milky-way-like ISM.
It is possible in rare circumstances that the host galaxy could produce the observed level of scattering if it was extremely turbulent or that the burst passed through extensive amounts of host galaxy ISM.
This indicates that the galaxies could have more turbulent ISM, the bursts are produced in exceptional local environments, or that the bursts are scattered by intervening extragalactic media such as an intervening galaxy halo or intracluster medium. 



The most recent seven ASKAP-detected FRBs in this sample were observed in an incoherent sum mode with voltage capture, which enabled host-galaxy identification. The localisation provides information of the host galaxy environment which will help describe the host galaxy ISM component and estimate the host galaxy DM. 
The high time resolution dynamic spectrum enabled by forming finer channelisation or coherently dedispersing the voltage time series will remove DM temporal smearing, revealing microsecond timescale structure.

In the future, further high time resolution observations of FRBs with host galaxy and host environment properties,  will enable detailed studies  the scattering-DM distribution of FRBs, and the properties of turbulent media along gigaparsec-length lines of sight. 
\section*{Data Availability}
No new data were generated or analysed in support of this research.
\section*{Acknowledgements}
We thank Devansh Agarwal for providing the observational data for FRB 180417.
HQ acknowledges the support of 
the Hunstead Merit Award for Astrophysics from 
the late Distinguished Professor Richard Hunstead (1943$-$2020) at the University of Sydney.
RMS, JPM and KB acknowledges the support of the Australian Research Council through grant DP180100857.
RMS also acknowledges support from the Australia Research Council through FT190100155. 
ATD acknowledges the support of the Australian Research Council through grant FT150100415.
TM acknowledges the support of the Australian Research Council through grant FT150100099.
The Australian SKA Pathfinder is part of the Australia Telescope National Facility which is managed by CSIRO. Operation of ASKAP is funded by the Australian Government with support from the National Collaborative Research Infrastructure Strategy. ASKAP uses the resources of the Pawsey Supercomputing Centre. Establishment of ASKAP, the Murchison Radio-astronomy
Observatory and the Pawsey Supercomputing Centre are initiatives of the Australian Government, with support from the Government of Western Australia and the Science and Industry Endowment Fund. We acknowledge the Wajarri Yamatji people as the traditional
owners of the Observatory site.




\bibliographystyle{mnras}
\bibliography{export-bibtex} 



\appendix


\bsp	
\label{lastpage}
\end{document}

%% file: scatfit.tex
170107 & $608.38_{-0.82}^{+0.93}$& 571
 & $\mathit{0.98_{-0.35}^{+0.30}}$
 & ${0.79_{-0.45}^{+0.37}}$
 & ${< 1.31}$&$-8.8$ \\
170416 & $523.65_{-0.47}^{+0.50}$& 484
 & $\mathit{2.45_{-0.31}^{+0.37}}$
 & ${2.20_{-0.42}^{+0.39}}$
 & ${< 1.41}$&$-7.4$ \\
170428 & $992.45_{-0.78}^{+0.76}$& 952
 & $\mathit{<1.35}$
 & ${<1.38}$
 & ${< 1.44}$&$-10.5$ \\
170707 & $233.88_{-0.92}^{+0.87}$& 198
 & $\mathit{2.09_{-0.32}^{+0.37}}$
 & ${<0.89}$
 & ${< 0.71}$&$-33.6$ \\
170712 & $312.28_{-0.24}^{+0.30}$& 274
 & $\mathit{<0.80}$
 & ${<0.78}$
 & ${< 0.60}$&$-15.0$ \\
170906 & $389.15_{-0.88}^{+0.89}$& 350
 & $\mathit{0.99_{-0.34}^{+0.31}}$
 & ${0.82_{-0.29}^{+0.20}}$
 & ${< 0.54}$&$-8.7$ \\
171003 & $465.06_{-0.25}^{+0.23}$& 425
 & $\mathit{0.72_{-0.22}^{+0.19}}$
 & ${0.73_{-0.25}^{+0.19}}$
 & ${< 0.74}$&$-11.1$ \\
171004 & $303.78_{-0.35}^{+0.33}$& 265
 & $\mathit{0.85_{-0.29}^{+0.27}}$
 & ${0.65_{-0.36}^{+0.37}}$
 & ${< 1.12}$&$-18.4$ \\
171019{$^\dagger$} & $462.10_{-0.65}^{+0.74}$& 425
 & $\mathit{2.84_{-0.14}^{+0.15}}$
 & ${2.36_{-0.35}^{+0.37}}$
 & ${< 1.26}$&$-9.4$ \\
171020 & $114.01_{-0.11}^{+0.09}$& 76
 & $\mathit{<0.58}$
 & ${<0.53}$
 & ${< 0.39}$&$-8.3$ \\
171116 & $618.17_{-0.44}^{+0.46}$& 582
 & $\mathit{1.43_{-0.30}^{+0.31}}$
 & ${1.10_{-0.53}^{+0.43}}$
 & ${< 1.85}$&$-9.1$ \\
171213 & $158.42_{-0.07}^{+0.06}$& 122
 & $\mathit{<0.43}$
 & ${<0.47}$
 & ${< 0.35}$&$-6.3$ \\
171216 & $203.74_{-0.28}^{+0.32}$& 167
 & $\mathit{0.85_{-0.53}^{+0.37}}$
 & ${0.70_{-0.43}^{+0.45}}$
 & ${< 0.76}$&$-13.2$ \\
180110 & $714.03_{-0.27}^{+0.25}$& 675
 & ${4.73_{-0.21}^{+0.16}}$
 & $\mathit{<0.74}$
 & $\mathit{5.92_{-0.26}^{+0.27}}$ &$137.0$\\
180119 & $401.40_{-0.27}^{+0.26}$& 366
 & ${1.97_{-1.12}^{+0.60}}$
 & $\mathit{<0.56}$
 & $\mathit{1.94_{-0.33}^{+0.37}}$ &$1.4$\\
180128.0 & $441.34_{-0.44}^{+0.44}$& 409
 & $\mathit{1.27_{-0.26}^{+0.27}}$
 & ${0.72_{-0.41}^{+0.43}}$
 & ${< 1.72}$&$-5.0$ \\
180128.2 & $495.40_{-0.62}^{+0.61}$& 455
 & $\mathit{1.29_{-0.18}^{+0.19}}$
 & ${<0.77}$
 & ${< 2.95}$&$-1.8$ \\
180130 & $343.09_{-0.77}^{+0.76}$& 304
 & ${3.54_{-0.68}^{+0.82}}$
 & $\mathit{0.73_{-0.44}^{+0.56}}$
 & $\mathit{5.95_{-1.08}^{+1.33}}$ &$8.4$\\
180131 & $657.45_{-0.58}^{+0.53}$& 618
 & $\mathit{2.34_{-0.30}^{+0.32}}$
 & ${1.54_{-0.66}^{+0.67}}$
 & ${< 2.34}$&$-4.7$ \\
180212 & $167.56_{-0.11}^{+0.11}$& 137
 & $\mathit{<0.50}$
 & ${<0.45}$
 & ${< 0.28}$&$-10.0$ \\
180315 & $478.90_{-0.35}^{+0.35}$& 378
 & $\mathit{<0.89}$
 & ${<0.89}$
 & ${< 0.52}$&$-14.6$ \\
180324 & $429.73_{-0.35}^{+0.38}$& 366
 & ${2.72_{-0.46}^{+0.52}}$
 & $\mathit{<1.15}$
 & $\mathit{2.98_{-0.60}^{+0.62}}$ &$0.0$\\
180417 & $474.65_{-0.14}^{+0.15}$& 449
 & $\mathit{0.86_{-0.10}^{+0.10}}$
 & ${0.86_{-0.12}^{+0.11}}$
 & ${< 0.48}$&$-10.8$ \\
180430 & $264.11_{-0.04}^{+0.03}$& 95
 & $\mathit{<0.54}$
 & ${<0.58}$
 & ${< 0.24}$&$-12.2$ \\
180515 & $354.88_{-0.19}^{+0.18}$& 322
 & $\mathit{<0.76}$
 & ${<0.75}$
 & ${< 0.52}$&$-14.6$ \\
180525 & $387.49_{-0.14}^{+0.16}$& 357
 & ${2.24_{-0.09}^{+0.08}}$
 & $\mathit{1.65_{-0.15}^{+0.15}}$
 & $\mathit{1.30_{-0.21}^{+0.20}}$ &$0.8$\\
180924 & $362.38_{-0.15}^{+0.16}$& 322
 & $\mathit{0.66_{-0.13}^{+0.12}}$
 & ${<0.63}$
 & ${< 0.72}$&$-12.5$ \\
181112 & $588.76_{-0.20}^{+0.19}$& 549
 & $\mathit{<0.60}$
 & ${<0.58}$
 & ${< 0.55}$&$-8.7$ \\
190102 & $364.38_{-0.22}^{+0.20}$& 307
 & $\mathit{<0.67}$
 & ${<0.62}$
 & ${< 0.55}$&$-11.2$ \\
190608 & $339.48_{-0.51}^{+0.52}$& 302
 & $\mathit{3.45_{-0.29}^{+0.31}}$
 & ${2.22_{-0.64}^{+0.63}}$
 & ${< 3.58}$&$-6.2$ \\
190611.2 & $321.37_{-1.07}^{+0.96}$& 264
 & $\mathit{1.42_{-0.52}^{+0.45}}$
 & ${1.03_{-0.63}^{+0.61}}$
 & ${< 1.45}$&$-12.4$ \\
190711{$^\dagger$} & $590.49_{-0.79}^{+0.73}$& 534
 & $\mathit{3.47_{-0.24}^{+0.26}}$
 & ${3.32_{-0.35}^{+0.30}}$
 & ${< 1.12}$&$-8.2$ \\
190714 & $503.80_{-0.30}^{+0.30}$& 465
 & $\mathit{<0.88}$
 & ${<0.88}$
 & ${< 0.78}$&$-12.1$ \\

%% file: section4.tex

\begin{figure}
    \centering
	\includegraphics[trim={0 0 0 0 },clip,width=\columnwidth]{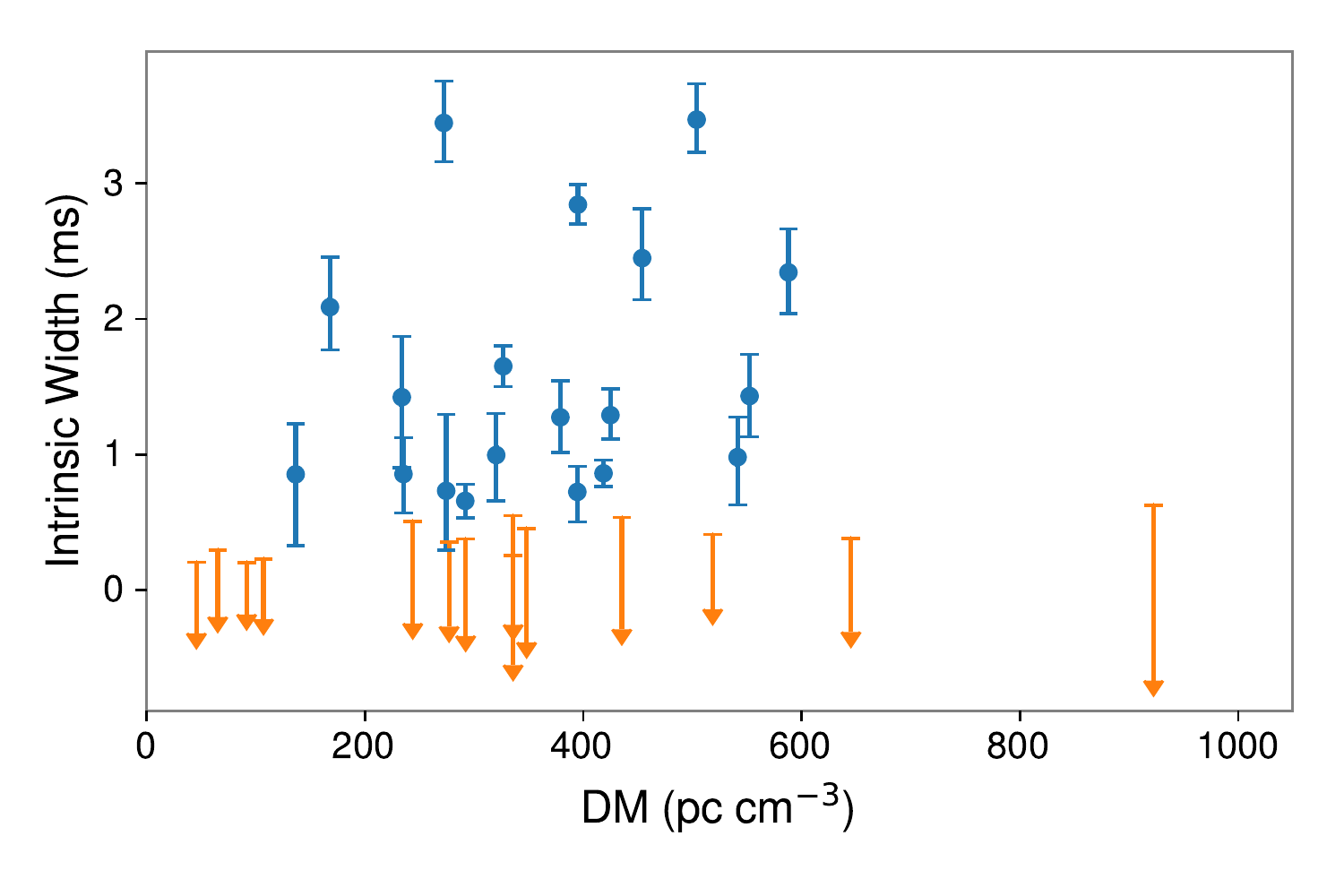}
    \\
    \includegraphics[trim={0cm 0cm 0 0cm },clip,width=1.0\columnwidth]{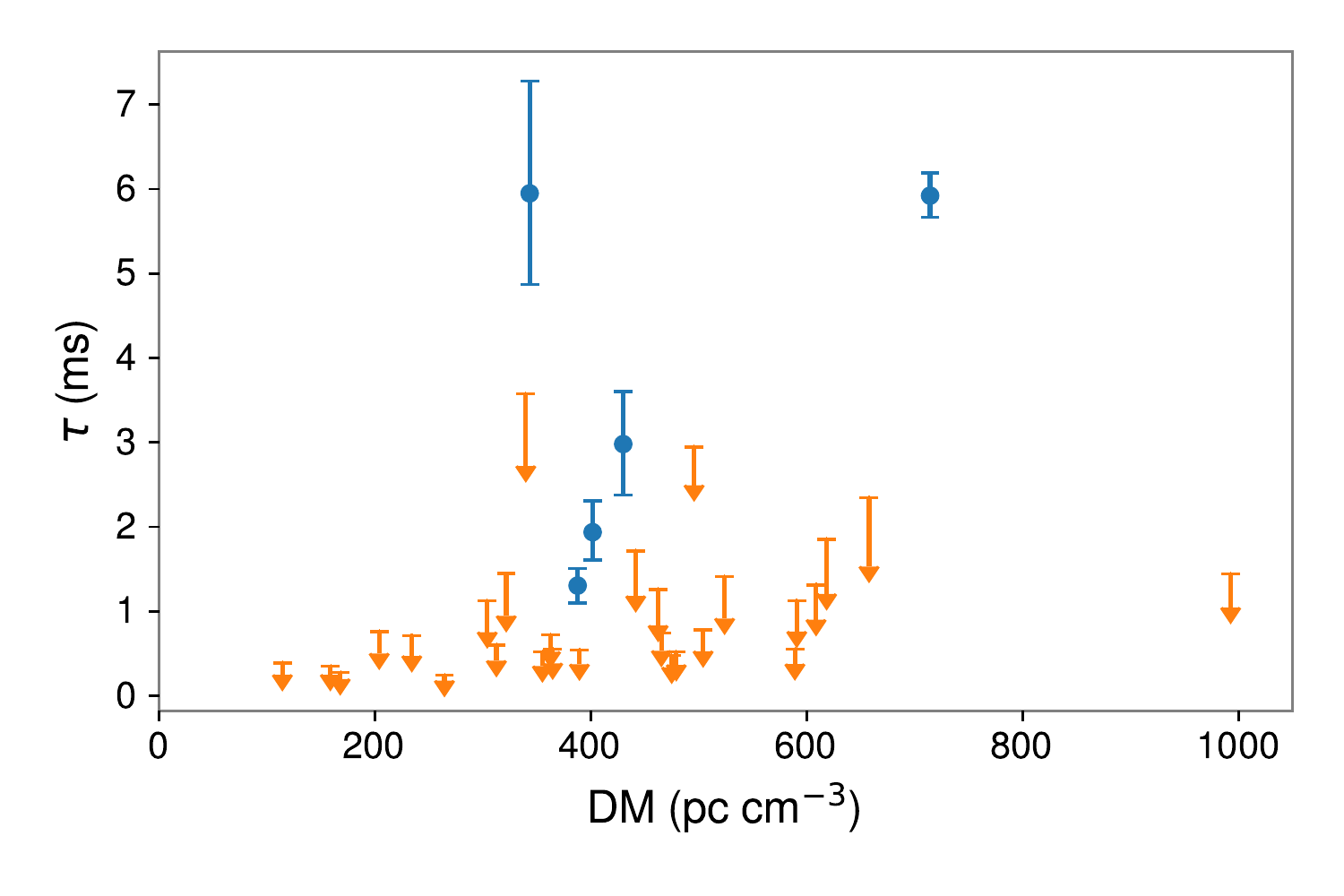}
    \\
	\includegraphics[trim={0cm 0cm 0 0cm },clip,width=1.0\columnwidth]{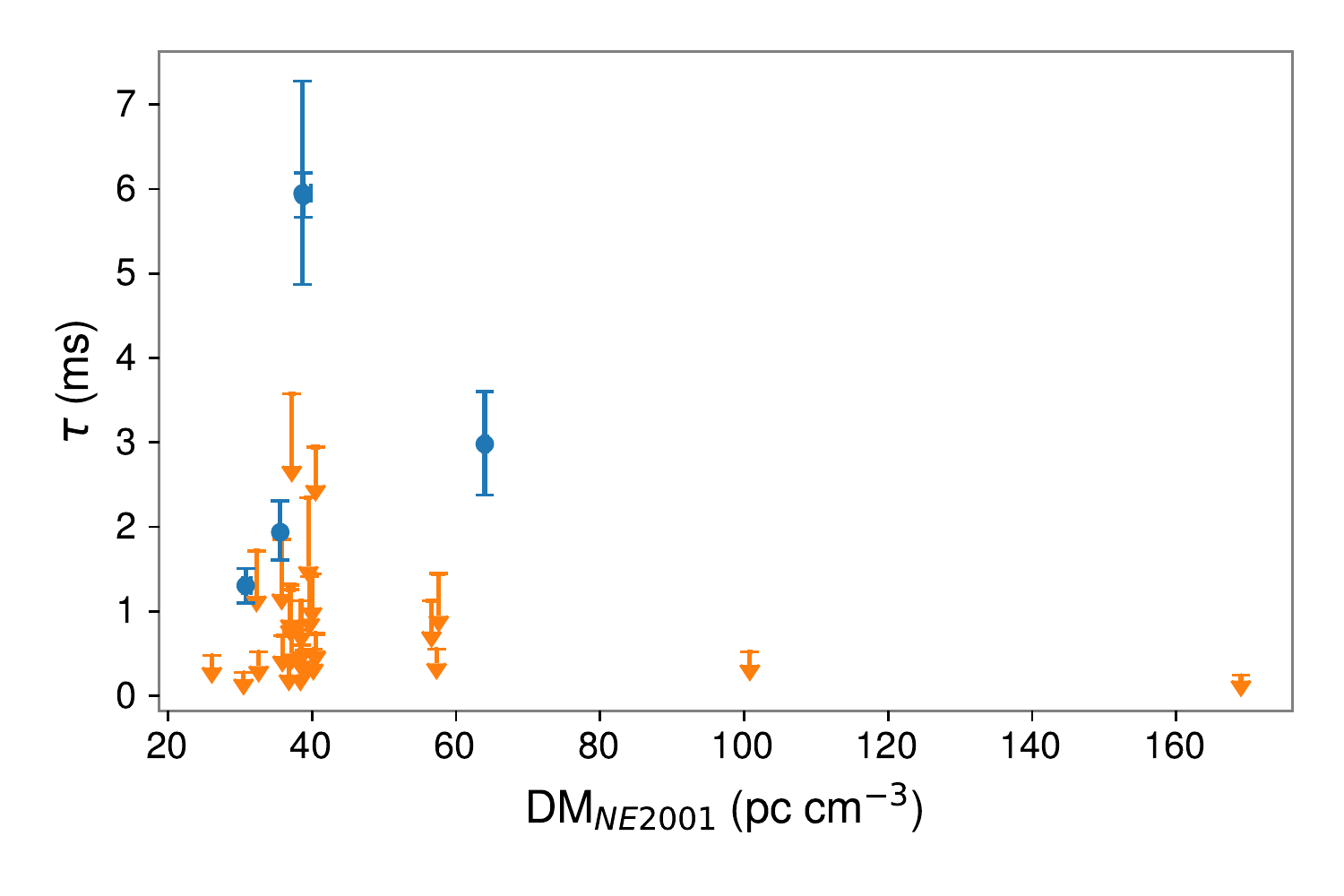}
    \caption{Distribution of FRB pulse measurements against DM, confirmed measurements are plotted in blue, upperlimits are plotted in orange.
    \textit{Top:}
    Intrinsic width measurement ($\sigma$) and DM of ASKAP FRBs.
    \textit{Middle:}Scattering time ($\tau$) of ASKAP FRBs plotted against total DM,
    showing no correlation between the total DM and pulse broadening.
    \textit{Bottom:}Scattering time ($\tau$) of ASKAP FRBs plotted over estimated line of sight Galactic DM contribution from the NE2001 electron density model ($\mathrm{DM_{NE2001}}$).}
    \label{fig:width_dm}
    \label{fig:scat_ism}
    \label{fig:scat_total}
\end{figure}


\begin{figure}
    \centering
	\includegraphics[trim={0 0 0 0 },clip,width=\columnwidth]{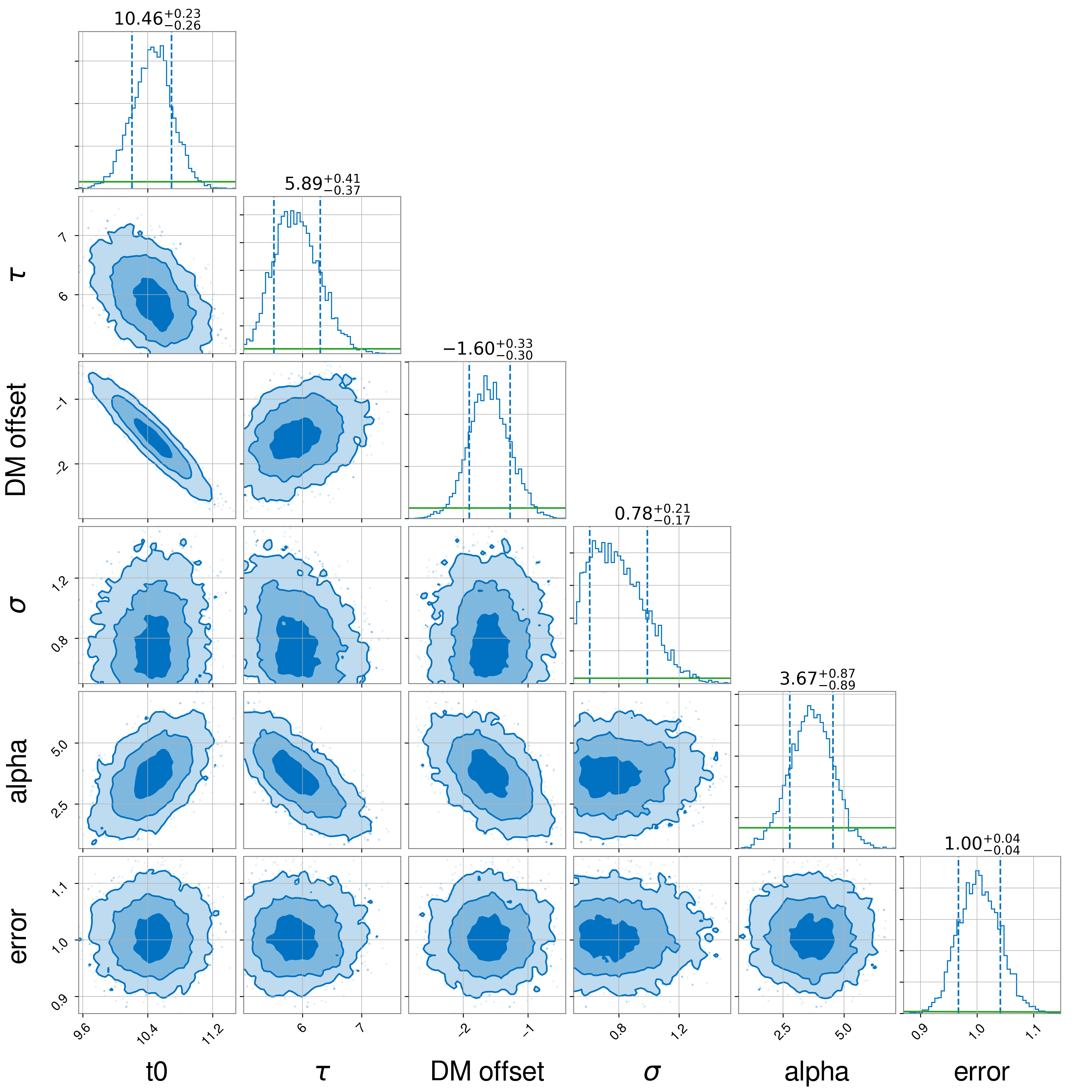}
    \caption{Posterior distribution of FRB 180110 using a scattering model with an unconstrained scattering index. 
    The parameters shown are the centre position of pulse (t0/ms), scattering timescale at 1.3 GHz ($\tau$/ms), DM$_{\mathrm{err}}$ (DM offset/\dmunit), pulse width ($\sigma$/ms),
    $\alpha$ (alpha) and uncertainty (error). Priors for each parameter is displayed as a green curve. The three shades of contour correspond to 1,2 and 3\ $\sigma$ confidence level.
    }
    
    \label{fig:post_180110}
\end{figure}


\begin{figure}
    \centering
	\includegraphics[trim={0.9cm 3cm 0 3cm },clip,width=1.1\columnwidth]{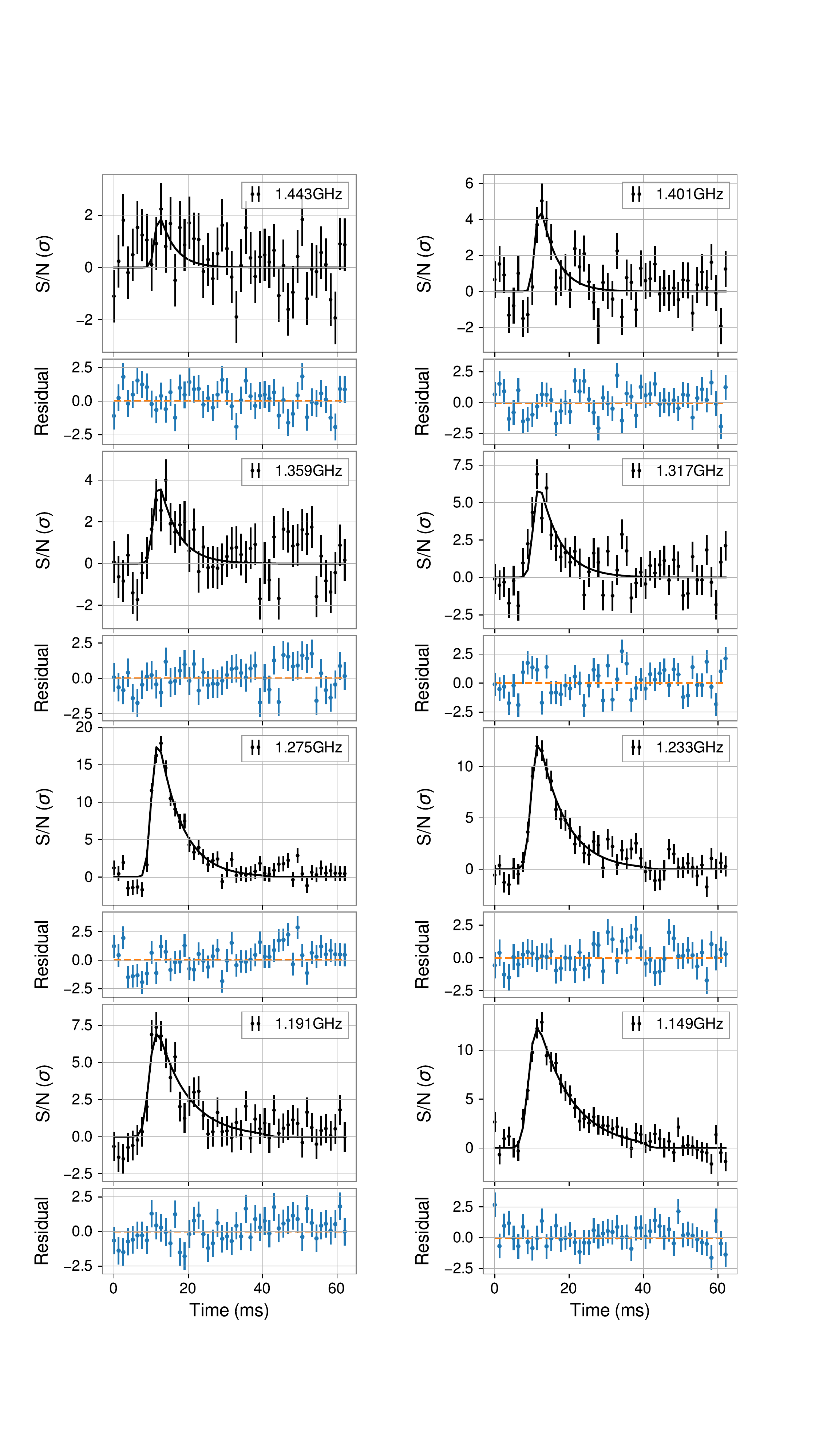}
    \caption{Subband pulse profiles of FRB 180110. We plot the best-fitting model and residual for each subband. The exponential broadening scales with frequency using an exponential power law with an index of $-3.8$.}
    \label{fig:residual_180110}
\end{figure}


\begin{figure}
    \centering
	\includegraphics[trim={0 0 0 0 },clip,width=\columnwidth]{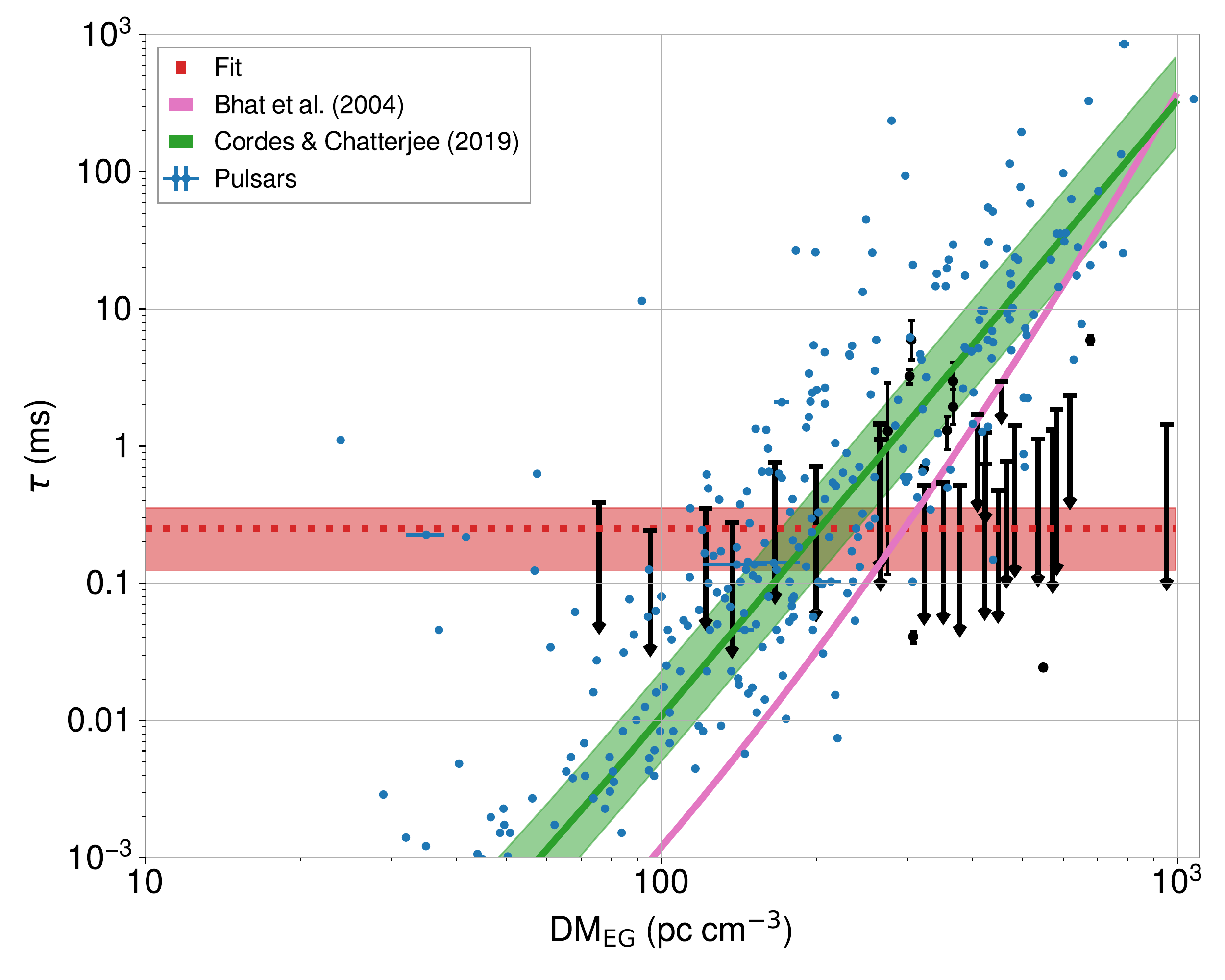}\\
    \caption{Pulse broadening measurement of ASKAP FRBs plotted against estimated extragalactic DM contribution. The scattering time, $\tau$ is scaled to 1.3 GHz with a power law index of --$4$. The population inference fit results from the flat model is plotted as an estimate limit of the scattering, the $1-\sigma$ upper limit and $2-\sigma$ lower limits is displayed as the shaded region in red. The scaled DM$-\tau$ relation from \citet{2019ARA&A..57..417C} is drawn with $1-\sigma$ uncertainty shaded region in green. We display the DM$-\tau$ relation of Galactic pulsars from \citet{2004ApJ...605..759B} and the measure of scattering from Galactic pulsars \citep{2005AJ....129.1993M} as comparison of the scattering expected from Milky-Way-like ISM.}
    \label{fig:scat_dm}
\end{figure}
